\documentstyle[12pt]{article}
\begin{document}
\title{Fractal Statistics}
\author{B.G. Sidharth$^*$\\
Centre for Applicable Mathematics \& Computer Sciences\\
B.M. Birla Science Centre, Adarsh Nagar, Hyderabad - 500 063 (India)}
\date{}
\maketitle
\footnotetext{$^*$Email:birlasc@hd1.vsnl.net.in; birlard@ap.nic.in}
\begin{abstract}
We consider the recent description of elementary particles in terms of Quantum
Mechanical Kerr-Newman Black Holes, a description which provides a rationale
for and at the same time reconciles the Bohm-hydrodynamical formulation on
the one hand and the Nelsonian stochastiic formulation on the other. The Boson-Fermion
divide is discussed, and it is pointed out that in special situations, anomalous
statistics, rather than Bose-Einstein or Fermi-Dirac states, can be encountered.
\end{abstract}
\section{Introduction}
In a recent model\cite{r1,r2} Fermions, in particular the electrons have
been treated as, what may be called Quantum Mechanical Kerr-Newman Black
Holes (QMKMBH), a treatment that leads to a successful interpretation of
several hitherto inexplicable features, both in Particle Physics and
Cosmology\cite{r3,r4,r5,r6}. In the hydrodynamical formulation these
QMKNBH can be considered to be vortices bounded by the Compton wavelength
in a description that leads to a harmonious convergence with Nelson's
stochastic theory also\cite{r7} (Cf.Discussion). In this picture, Bosons would be, not
vortices, but rather streamlines\cite{r8}.\\
In what follows, we shall show that within this framework it is possible
to explain the divide between Fermi-Dirac and Bose-Einstein statistics, as
also examine special situations where there would be fractal statistics,
that is Bosonisation of Fermions and Fermionisation of Bosons.
\section{Usual Statistics}
We first observe that the above vortex and streamline description provides
an explanation for the Fermionic and Bosonic statistics. Indeed, let $n_K$
be the occupation number for the energy or momentum state defined by $K$. For
Fermions $n_K = 0$ or $1$, where as $n_K$ can be arbitrary for Bosons. The
reason is that Fermions are bounded by the Compton wavelength. That is,
they are localised, a description which requires both negative and positive
energy solutions\cite{r9}, which infact is expressed by zitterbewegung
effects. The localisation in space automatically implies an indeterminacy
of energy or momentum $K$. Thus, though an energy or momentum state, in
practical terms implies a small spread $\Delta K$, it is not possible to
cram Fermions which also have a momentum energy interdeterminacy spread, arbitrarily
into this state.\\
On the other hand Bosons are not bound by the Compton wavelength vortices, and
so have sharper momentum states, so that any number of them can be crammed into
the state $K$ (which really is blurred by the Uncertainity Principle
indeterminacy of $\Delta K$). Another way of expressing these facts are by
saying that the Fermionic wave function in space is weak, but not the Bosonic
wave function\cite{r10}, the latter fact being symptomatic of a field or an
interaction.\\
The above considerations immediately follow from a recent description in
terms of quantized fractal space time\cite{r11}. In this case we have a
non commutative geometry given by
\begin{equation}
[x,y] = 0 (l^2), [x,p_x]=\imath \hbar [1+l^2]\label{e1}
\end{equation}
where $l$ is the Compton wavelength.
It is precisely the space quantization at the Compton scale that leads to the
Dirac matrices and their anti-commutation relations.\\
Bosons on the other hand
would be bound states of the Fermions (Cf.\cite{r12} and other references), or
alternatively they would be a super position of vortices leading to a streamline
like description\cite{r8}.
\section{Fractal Statistics}
However there could be certain special situations in which the above space
localised and momentum space localised description of Fermions and Bosons
gets blurred, in which case anomalous or fractal statistics would come into
play. This could happen, for example when the Compton wavelength $l$ of the
Fermion becomes very small, that is the particle is very massive. In this case
the non commutativity of the geometry referred to above in (\ref{e1}) disappears
and we return to the usual commutation relations of non relativistic Quantum
Mechanics, that is a description in terms of the spinless Schrodinger equation.
Indeed in this case, $v$ being the velocity of the particle $v/c$ would be
small and the Dirac equation tends to the Schrodinger equation\cite{r9}.\\
There would thus be a Bosonisation effect. This would also be expected at very
low temperatures, for example below the Fermi temperature, when the energy spread
of the Fermions would itself be small, rather as in the case of Bosons and
anomalous behaviour, for example on the lines of the superfluidity of
$He^3$\cite{r13} can be expected.\\
For very light Fermions, for example Neutrinos, the Compton wavelength be very
large, but in this case the double connectivity of the QMKNBH disappears and the
observed anomalous features of the Neutrino show up, as discussed
elsewhere\cite{r1,r14}.\\
Indeed with recent developments in nano technology and thin films, we are able to
consider one dimensional and two dimensional Fermions, in which case, Bosonisation
effects show up as discussed elsewhere\cite{r13}. In any case in the one
dimensional and two dimensional cases, the Dirac equation becomes a two
component equation, without an invariant mass\cite{r15}, while at the same time,
handedness shows up\cite{r16}. It is worth mentioning here that the Dirac
matrices can have only even dimensionality, corresponding to the above
anomalous two component Dirac spinors, and the usual Dirac bispinors of the
three dimensional theory.\\
We now consider in greater detail two illustrative
situations where anomalous behaviour shows up.\\
{\bf \large 1. Nearly Mono Energetic Fermions (Cf.\cite{r17}):}\\
Our starting point is the well known formula for the occupation number of
a Fermion gas\cite{r18}
\begin{equation}
\bar n_p = \frac{1}{z^{-1}e^{bE_p}+1}\label{e2}
\end{equation}
where, $z' \equiv \frac{\lambda^3}{v} \equiv \mu z \approx z$ because, here,
$\mu \approx 1$ (Cf. Appendix);
$$v = \frac{V}{N}, \lambda = \sqrt{\frac{2\pi \hbar^2}{m/b}}$$
\begin{equation}
b \equiv \left(\frac{1}{KT}\right), \quad \mbox{and} \quad \sum \bar n_p = N\label{e3}
\end{equation}
Let us consider particular a collection of Fermions which is somehow made
nearly mono-energetic, that is, given by the distribution,
\begin{equation}
n'_p = \delta (p - p_0)\bar n_p\label{e4}
\end{equation}
where $\bar n_p$ is given by (\ref{e2}).\\
This is not possible in general - here we consider a special hypothetical situation
of a collection of mono-energetic particles in equilibrium which is the idealization
of a contrived experimental set up. For example, the following is a mono
energetic equilibrium distribution function:
$$f \alpha exp [-\rho (|\vec v | - | \vec v_0 |)^2], \rho \equiv m/2KT > > 1,$$
Infact one can show that the above function is consistent with the Bolitzmann
equation for a collection of particles under a uniform magnetic field $\vec B$:
Infact
$$\frac{\partial f} {\partial t} = 0,$$
while one can easily show that
$$\vec \Delta f = 0, \vec \Delta_{\vec v} [exp[-\rho (|\vec V | - |\vec V_0 |)^2]
\alpha [(|\vec V | - |\vec V_0 |) f\vec V ]$$
Further the force is given by
$$\vec F \alpha \vec V \times \vec B$$
so that the force term in the Boltzmann equation viz.,
$$\frac{\vec F}{m} \cdot \vec \Delta_{\vec v} f \alpha \vec V \times \vec B \cdot
\vec V = 0$$
Finally\\
$\frac{\partial f}{\partial t} = 0$ means $\frac{dH}{dt} = 0$ that is
$(\frac{\partial f}{\partial t})_{coll} = 0$\\
Thus the above distribution function which becomes mono energetic for $\lambda > > 1$
satisfies the Boltzmann equation (Cf.ref.\cite{r18}); when $\rho \to \infty$
we get the above $n'_p$ in this case.\\
By the usual formulation we have,
\begin{equation}
N = \frac{V}{\hbar^3} \int d\vec p n'_p = \frac{V}{\hbar^3} \int \delta (p - p_0)
4\pi p^2\bar n_p dp = \frac{4\pi V}{\hbar^3} p^2_0 \frac{1}{z^{-1}e^{\theta}+1}\label{e5}
\end{equation}
where $\theta \equiv bE_{p_0}$.\\
It must be noted that in (\ref{e5}) there is a loss of dimension in momentum
space, due to the $\delta$ function in (\ref{e4}) - infact such a fractal two
dimensional situation would in the relativistic case lead us back to the anomalous
behaviour already alluded to\cite{r19}. In the non relativistic case two
dimensions would imply that the coordinate $\psi$ of the spherical polar
coordinates $(r, \psi , \phi)$ would become constant, $\pi /2$ infact. In
this case the usual Quantum numbers $l$ and $m$ of the spherical harmonics\cite{r20}
no longer play a role in the usual radial wave equation
\begin{equation}
\frac{d^2u}{dr^2} + \left\{\frac{2m}{\hbar^2}[E - V(r)] - \frac{l(l+1)}{r^2}\right\}
u = 0,\label{e6}
\end{equation}
The coefficient of the centrifugal term $l(l+1)$ in (\ref{e6}) is replaced by $m^2$ as in
Classical Theory\cite{r21}.\\
To proceed, in this case, $KT = <E_p> \approx E_{p)}$, so that, $\theta \approx 1$.
But we can continue without giving $\theta$ any specific value.\\
Using the expressions for $v$ and $z$ given in (\ref{e3}) in (\ref{e4}), we get
$$(z^{-1} e^\theta + 1) = (4\pi )^{5/2} \frac{z^{'-1}}{p_0};\mbox{whence}$$
\begin{equation}
z^{'-1}A\equiv z^{'-1}\left(\frac{(4\pi )^{5/2}}{p_0} - e^\theta\right) = 1,\label{e7}
\end{equation}
where we use the fact that in (\ref{e3}), $\mu \approx 1$ (Cf.Appendix).\\
A number of conclusions can be drawn from (\ref{e7}). For example, if,
$$A \approx 1, i.e.,$$
\begin{equation}
p_0 \approx \frac{(4\pi )^{5/2}}{1+e}\label{e8}
\end{equation}
where $A$ is given in (\ref{e7}), then $z' \approx 1$. Remembering that in
(\ref{e3}), $\lambda$ is of the order of the de Broglie wave length and $v$
is the average volume occupied per particle, this means that the gas gets very
densely packed for momenta given by (\ref{e8}). Infact for a Bose gas, as is
well known, this is the condition for Bose-Einstein condensation at the level
$p = 0$ (cf.ref.\cite{r18}).\\
On the other hand, if,
$$A \approx 0 (\mbox{that}\quad \mbox{is}\quad \frac{(4\pi )^{5/2}}{e} \approx
p_0)$$
then $z' \approx 0$. That is, the gas becomes dilute, or $V$ increases.\\
More generally, Equation (\ref{e7}) also puts a restriction on the energy (or
momentum), because $z' > 0$, viz.,
$$A > 0(i.e.p_0 < \frac{(4\pi )^{5/2}}{e})$$
$$\mbox{But \quad if}A < 0, (i.e.p_0 > \frac{(4\pi )^{5/2}}{e})$$
then there is an apparent contradiction.\\
The contradiction disappears if we realize that $A \approx 0$, or
\begin{equation}
p_0 = \frac{(4\pi )^{5/2}}{e}\label{e9}
\end{equation}
(corresponding to a temperature given by $KT = \frac{p^2_0}{2m}$) is a threshold
momentum (phase transition). For momenta greater than the threshold given
by (\ref{e9}), the collection of Fermions behaves like Bosons. In this case,
the occupation number is given by
$$\bar n_p = \frac{1}{z^{-1}e^{bE_p}-1},$$
instead of (\ref{e2}), and the right side equation of (\ref{e7}) would be given
by $' -1'$ instead of $+1$, so that there would be no contradiction. Thus in
this case there is an anomalous behaviour of the Fermions.\\
The Bosonic behaviour of Fermions can be understood in a simple way from a
different standpoint. Let us consider the case described before equation (\ref{e5}),
of a collection of Fermions with $m > > 1$, and consequently all having nearly
the same momentum $\vec p$. In this case, it is possible to choose a Lorentz
frame in which all the particles are nearly at rest, i.e.,$v/c \approx 0$. As mentioned, it is known that the Dirac equation describing spin $1/2$ particles
goes over into the Schrodinger equation describing spinless particles\cite{r9}.
Effectively, we have a collection of Bosons.\\ \\
{\bf \large 2. Degenerate Bosons:}\\
We could consider a similar situation for Bosons also (Cf.\cite{r22} where an
equation like (\ref{e4}) holds. In this case we have equations like (\ref{e8})
and (\ref{e9}):
\begin{equation}
p_0 \approx \frac{(4\pi )^{5/2}}{1.4e-1}\label{e10}
\end{equation}
\begin{equation}
p_0 \approx \frac{(4\pi )^{5/2}}{e}\label{e11}
\end{equation}
((\ref{e11}) is the same as (\ref{e9}), quite expectedly).
At the momentum given by (\ref{e10}) we have a densely packed Boson gas
rather as in the case of Bose Einstein condensation. On the other hand at the
momentum given by (\ref{e11}) we have infinite dilution, while at lower
momenta than in (\ref{e11}) there is an anomalous Fermionisation.\\
Finally it may be pointed out that at very high temperatures, once again the
energy - momentum spread of a Bosonic gas becomes large, and Fermionisation
can be expected, as in indeed has been shown elsewhere\cite{r23}. In any
case at these very high temperatures, we approach the Classical Maxwell
Boltzmann situation.
\section{Discussion}
1. It was mentioned in Section 1 that in the QMKNBH description there is a
reconciliation between Nelson's stochastic theory and the Bohm-Hydrodynamical
approach, once it is realized that the diffusion constant of Nelson's theory
is related to the Compton wavelength, and that the non local Bohm potential
gives the energy of the particle (Cf.\cite{r1,r7} for details). Infact in the
Nelsonian theory, space time is a non differenciable manifold, there being a
Double Weiner process which leads to the usual solinoidal velocity given by
$\vec v = \frac{\vec \nabla S}{m}$ as also an osmotic velocity given by
$\vec u = \frac{\hbar}{2m} \frac{\vec \nabla R}{R}$ where the Quantum Mechanical
wave function $\psi = Re^{\frac{\imath}{\hbar}}S.$ We get identical expressions in the
Bohm-hydrodynamical approach also. Indeed if there were no double Weiner process,
then in the Nelsonian theory $\vec u$ would vanish and so would $\nabla^2 R$ and
$Q$, that is the Compton wavelength vortex and the mass and the energy of the
particle would both disappear.\\
2. Interestingly in the QMKNBH-Hydrodynamical vortex picture as in the usual
Quantum Theory of addition of angular momentum, we can recover the fact that
the sum or bound state of two such vortices or spin half particles would
indeed give Bosons.\\
This can be seen as follows, from the theory of vortices\cite{r24}.
The velocity distribution is given by
\begin{equation}
v = \Gamma /2\pi r\label{e12}
\end{equation}
In the case of the QMKNBH vortex we have to use in (\ref{e12}) $v = c \quad \mbox{and}
\quad r = \hbar/2mc$, the Compton wavelength of the particle. So we have $\Gamma
= \frac{h}{m}$ which is also the diffusion constant of Nelsonian Theory.\\
If we consider two parallel spinning vortices separated by a distance $d$, then
the angular velocity is given by
$$\omega = \frac{\Gamma}{\pi d^2}$$
whence the spin of the system turns out to be $h$, that is in usual units the
spin is (\ref{e1}), wth states $\pm 1$.\\
There is also the case where the two above vortices are anti parallel. In this
case there is no spin, but rather there is the linear velocity given by
$$v = \Gamma/2 \pi d$$
This corresponds to the spin $1$ case with state $0$.\\
Together, the two above cases give the $3,-1, 0, +1$ states of spin $1$ as in the
Quantum Mechanical Theory.\\
It must be noteed that the distance $d$ between vortices could be much greater
than the Compton wavelength scale, so that the wave function of the Boson in
the above description would be extended in space in comparison to the Fermionic
wave function, as pointed out in the text.

\end{document}